\documentclass[a4paper,twoside]{article}
\newcommand{\affil}[1]{$^{\rm #1}$}
\usepackage[authoryear]{natbib}
\bibpunct{(}{)}{;}{a}{}{,}
\usepackage{graphicx}
\usepackage{rotating}
\usepackage{textcomp}
\usepackage{array}
\date{} 
\title{\large\bf\flushleft Testing Potential New Sites for Optical Telescopes in Australia}
\author{\parbox{\textwidth}{\flushleft
\vspace{-0.5cm}
{\it C.E. Hotan\affil{A,B}, S.J. Tingay\affil{A}, 
and K. Glazebrook\affil{C}}\\
\vspace{0.4cm}
{\small \affil{A}\,ICRAR - Curtin University, GPO Box U1987, Perth, WA 6845}\\
{\small \affil{B}\,Email: cehotan@gmail.com}\\
{\small \affil{C}\,Swinburne University of Technology, PO Box 218, Hawthorn, Vic 3122}}}
\begin{document}
\twocolumn[
\begin{changemargin}{.8cm}{.5cm}
\begin{minipage}{.9\textwidth}
\vspace{-1cm}
\maketitle
%
%
\small{\bf Abstract:} 
In coming years, Australia may find the need to build 
new optical telescopes to continue local programmes, contribute to global survey projects, and form a local multi-wavelength connection for the new radio telescopes being built. In this study, we refine possible locations for a new optical telescope by studying remotely sensed meteorological infrared data to ascertain expected cloud coverage rates across Australia, and combine these data with a Digital Elevation Model using a Geographic Information System. We find that the best sites within Australia for building optical telescopes are likely to be on the highest mountains in the Hamersley Range in Northwest Western Australia, while the MacDonnell Ranges in the Northern Territory may also be appropriate. We believe that similar seeing values to Siding Spring should be obtainable and with significantly more observing time at the identified sites. We expect to find twice as many clear nights 
as at current telescope sites. 
These sites are thus prime locations for future on-site testing.

Accepted for publication in PASA.

\medskip
\medskip
\end{minipage}
\end{changemargin}
]
\small

\section{Introduction}
In this paper we use meteorological and geographic analyses to identify locations in Australia which may be capable of hosting significant optical astronomy research facilities, potentially 
up to the 8m class of telescopes.\\
We wish to ascertain where in Australia ``good'' locations for building optical telescopes are, based on the principles that a good site has a high elevation, little atmospheric distortion, and good ``seeing'', with minimal cloud cover reducing available observing time. Informally it is thought that locations may exist in Australia with overall more favourable conditions than those found at Siding Spring. \\

Australia is a continent with vast flat regions and relatively few mountain ranges. Those mountains that do exist are typically not of very high altitude. Nevertheless, Australia has an active astronomical research community in a range of wavelengths, including optical astronomy through collaborations on larger telescopes overseas, and advanced instrument engineering programmes at the AAO. Optical astronomy has a need for a range of telescope sizes. New small telescopes are often developed to test instruments or execute new science concepts (for example the SkyMapper telescope \citep{SkyMapper}), and it is convenient to have them in geographic proximity.\\ 

The major optical research site in Australia is at Siding Spring near Coonabarabran in northern New South Wales. This site hosts the Australian Astronomical Observatory (AAO) and the Mount Stromlo and Siding Spring Observatory (MSSSO). Other research telescopes are located at Mt Stromlo, A.C.T., Canopus Hill, Tas. (soon to move to Bisdee Tier), Bickley, W.A. and Gingin, W.A.\\ 

``Seeing'' is a measurement used by astronomers to quantify the stability of the atmosphere through which they are observing. In general, higher altitude is correlated with better seeing as one is looking through less turbulent layers in the atmosphere. 
Construction of a new telescope requires a site with a high altitude and clear, stable atmosphere and very little artificial light pollution.\\ 

An unpublished analysis of 1999 infrared and precipitable water vapour data by KG (private communication) suggested Mt Bruce, W.A. as such a potential site. Based on similar analyses, \citet{Wood} had previously proposed Freeling Heights, S.A., which is near the Mount Searle site which was tested in conjunction with Siding Spring by \citet{Hogg65}. \citet{Walsh} suggests Mt Singleton in W.A. may also have comparable seeing to both Siding Spring and Freeling Heights. This paper aims to quantitatively further these investigations by conducting a study of Australian continental cloud cover, altitude, and a number of other geographic considerations, to identify all possible locations for future research optical telescopes in Australia.\\

As large telescopes (of any wavelength) require significant capital expenditure, it is vital to choose the right location. Any telescope will have some down­time both due to breaks for maintenance, and sky conditions including clarity of atmosphere and operability of telescope given wind and weather.\\

With a diameter of 3.9m, the Australian Astronomical Telescope at the AAO at Siding Spring is the largest optical astronomy research instrument in Australia\footnote{http://www.aao.gov.au/AAO/about/aat.html},
however it is closed due to weather on average around 35-­40\% of the (night) time, with “photometric” (ideal clear sky) conditions for only 40\% of the remaining observable time \citep{Tinney, Lee}.\\ 

Geographic and meteorological studies into telescope siting have been done in the past \citep{Ardeberg, Coops, Graham, Hogg65, Zhu}. Historically, such studies were temporally limited. In terms of Australian optical astronomy, a description of the AAO site selection process is found in \citet{Gascoigne}. In recent years, the University of Friboug, Switzerland, has created its own Geographic Information System for the purpose of telescope siting questions, designed with the motivation of selecting a site for the proposed European Extremely Large Telescope (E-ELT) in mind \citep{Graham, Sarazin06}.\\ 

Any proposed telescope might be of particular interest for multi-wavelength investigations with new radio astronomy facilities being built in Australia such as ASKAP \citep{Johnston}, the MWA \citep{Lonsdale} and potentially the SKA \citep{Dewdney}.\\

The research method employed in this project involves a low-resolution Geographic Information System style Multi­-Criteria Decision Analysis \citep{MCDAsotas, Longley, Malc} performed across Australia to identify candidate regions, using data freely available from the Bureau of Meteorology\footnote{The Australian Gov't Bureau of Meteorology (BoM) Geostationary Satellite Data Archive, \\http://www.bom.gov.au/sat/archive\_new/gms/} and Geoscience Australia\footnote{Geoscience Australia free data downloads, \\https://www.ga.gov.au/products/}.
The meteorological data were initially analysed using basic image processing code written by CH \citep{Hotan}, using Python. A brief description of this process is found in Section 2 of this paper.\\
Multi-Criteria Decision Analysis (MCDA) is performed using industry standard software, ESRI ArcGIS v9.3, into which the Geoscience Australia data can be readily loaded and analysed, with the meteorological data being rectified into a standard projection as appropriate in Section 3. Once all relevant datasets are present and correctly projected they can be added together to show the effects of different ``suitability metrics''.

\section{Meteorological Analysis}
\subsection{Why meteorology is important in choosing a telescope site}
We consider cloud cover as a proxy for the amount of observing time available at a site.\\ 

A number of variables must be taken into account when considering the siting for a telescope, as discussed by \citet{Ardeberg} wherein the author creates a list of 19 parameters in 4 categories which must be considered when siting a telescope. The first category is characteristics of the atmosphere.

\subsection{Directly sampled data}                                                                                                                                                                              In this investigation we have used only data freely available for download from the Australian Government Bureau of Meteorology (``BoM''). Long term data archives are only available on request and for a fee. In this study we limit ourselves to data which can be freely obtained.\\

The time period for which data were acquired in this investigation was June 2008 until January 2010. For the entire duration of this period, the Japan Meteorological Agency's MTSAT--1R has been the primary data source provided by the BoM. As all data come from this one source, our data are readily comparable with images taken at the same waveband. 

\subsubsection{The advantages of infrared data}
We have chosen to work with infrared (IR) wavelength data, in the $10.3-11.3\mu m$ waveband. This wavelength is sensitive particularly to cold atmospheric responses, and this is a good tracer of cloud cover, in particular high altitude cloud as well as rain--bearing cloud. 
This helps gauge whether conditions might be photometric. 
An advantage of using IR data is that these wavelengths can return readings at any time of day, in spite of some diurnal variations.\\

\subsubsection{Types of data}
Meteorological data may be acquired in a number of ways, remotely sensed from space, by sensors deployed at stations in various locations, or directly observed by eye. We have used satellite data as they give even coverage across the continent.\\

There is no fundamental reason to limit this study to directly sensed data from satellites. Derived data products and data observed at stations on the ground are also very important, and if available, should be taken into consideration. We have not done so here as BoM observation points are sparse in regions of interest. 
Derived information from satellite data, and interpolation of ground-based data, can permit the production of other potentially useful metrics such as precipitation, and wind strength and direction. A range of these products are freely available from the BoM\footnote{A range of climate data maps can be accessed on http://www.bom.gov.au/climate/averages/maps.shtml}.\\

\textit{Rainfall}
is of some interest to us in terms of seasonal variations. In Australia, rainfall is concentrated on the East coast, tropics, and southwest corner, falling predominantly in summer months in the tropics but in winter months at the lower latitudes.\\

\textit{Wind} 
data are available from the BoM. We may consider the wind across a large area, or we may be interested in wind patterns at a certain point. Wind data have not been incorporated in this study, but are discussed further in \citet{Hotan}. 
From an astronomical perspective, a site does not need to be especially calm for an optical telescope. A moving air column with constant direction and low speed may produce better seeing than a still air column that is prone to diurnal convection currents, or a turbulent air current. Laminar flow is often found across mountains in western coastal areas of continents, and when available, provides an important advantage in astronomical seeing \citep{McInnes}.\\ 

No indicative values for wind speed and direction or rain fall are available at our candidate sites from BoM. Information about wind could be obtained from global reanalyses such as \citet{NCFSR} NCFSR, or Atmosphere--Ocean coupled General Circulation Models, but was considered beyond the scope of this project.

\subsection{Acquisition and analysis of meteorological data}
The Bureau of Meteorology hosts a website from which images may be downloaded free of charge. However these images are of a poor resolution, roughly 30km per pixel, and being lossy JPEG images, suffer from aliasing effects from the superimposed country outlines map. Higher resolution images, roughly 10km per pixel, from the same satellite are also available, having an archive history of 20 days only. Thus it was necessary to download satellite images via FTP approximately every 2 weeks for a period of slightly over 18 months covering the period $11^{th}$ June 2008 to $22^{nd}$ January 2010.\\

To investigate cloud cover we wish to know where the cloud is in each satellite image according to pixel value. Images can be queried and manipulated in a number of ways. We used the functionality of the Image module from the Python Image Library (PIL) to open and interpret the data files and manipulate them in custom scripts. More information about the methods and algorithms used to combine the satellite images can be found in \citet{Hotan}.

\subsection{Results of meteorological analysis}
The images are greyscale 8--bit images. Every pixel has cloud in it some of the time over our period of interest, and as we have not applied thresholding to our data, any changes due to land temperature have not been excluded.\\ 

\begin{figure}[htbp]
	\centering
		\includegraphics[scale=0.25]{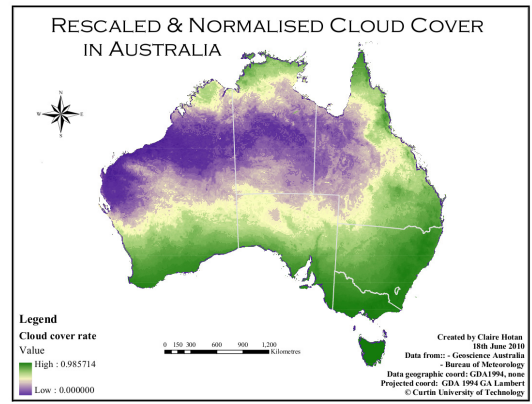}
		\caption{Map showing relative cloud cover across Australia at false {\raise.17ex\hbox{$\scriptstyle\sim$}}10km resolution in the period $11^{th}$ June 2008 - $22^{nd}$ January 2010.}
	\label{fig:cloudMap}
\end{figure}

For our GIS analysis we work with the combined image covering the full period for which we have data. Ideally we should have an integer number of years with equal days, nights, summers and winters, but in practice we have approximately 18 months with approximately 1.5 summers and winters due to the time constraints of the project. 
A rescaled version of this image, loaded into a GIS and recoloured, is shown in Figure \ref{fig:cloudMap}, where pixel value corresponds to the value in the IR images averaged over the period acquired. We show below that this is reasonably proportional to the number of cloudy days at a site.\\

From inspection of the total cloud cover image we see that there are apparently promising regions of low cloud cover around northwest Western Australia and extending inland toward the Northern Territory. By eye there appears to be some correlation between altitude and cloud cover. Suitably elevated sites in the Hamersley Range in W.A., such as Mount Bruce and Mount Meharry (both {\raise.17ex\hbox{$\scriptstyle\sim$}}1200m elevation), may be good candidate sites, in agreement with KG's analysis and the work of \citet{Coops}. \citet{McInnes} also mention the potential suitability of northern Western Australia.\\
We also suggest that an appropriate site in the MacDonnell Ranges in the N.T. would be a good candidate site (with areas over 1000m elevation). With no special knowledge of the area, a site was selected to give typical values for these Ranges, at $-23.89^oS$, $132.20^oE$.\\ 
\citet{Coops} found two appropriate sites based on their cloud cover analysis, the Hamersley Range in W.A., and the northern end of the Flinders Ranges in South Australia (previously tested by \citet{Hogg65}, and later tested by \citet{Wood}). While these sites are reputed to have very good seeing our cloud summation does not appear to be in agreement with this suggestion.\\

\begin{figure}[htbp]
	\centering
		\includegraphics[scale=0.3,angle=270]{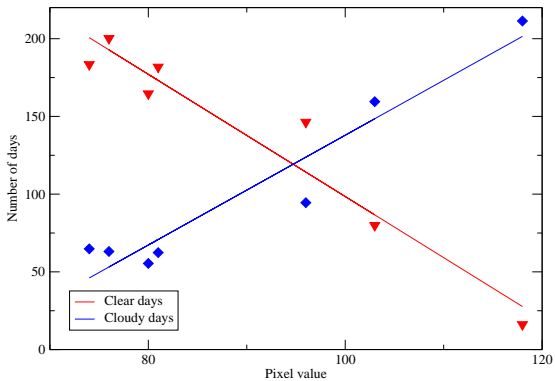}
		\caption{Plot of number of clear and cloudy days per year at selected BoM sites.}
	\label{fig:ClearCloudy}
\end{figure}

We are interested in the range of pixel values in our combined image, and how they relate to predicted available observing time. To quantify this we select 7 sites widely distributed around Australia which we believe are identifiable and near enough to weather stations that we can assign a pixel value to that station. These should form a reasonable basis for calibration. Stations selected have historical data of varying lengths, but covering the period of our satellite imagery. The BoM classify days as ``clear'' or ``cloudy''. These clear and cloudy day values for each identified site are shown in Table \ref{tab:ClearCloudy}, and plotted in Figure \ref{fig:ClearCloudy}. It appears that there is an approximately linear relationship between the number of clear (or cloudy) days and pixel value in the range 50-120. 
This validates our use of pixel value as a proxy for cloud cover, meaning that for any given site around Australia, simply measuring the pixel value for that location should give a reasonable estimate of annual cloud cover at that site. 
In this period, the AAO experienced typical observing conditions \citep{Lee}. Table \ref{tab:ClearCloudy} also shows the number of clear days we would predict based on linear regression of Figure \ref{fig:ClearCloudy} for each site.\\

\begin{table*}[htbp]
	\centering
		\begin{tabular}{|c|c|c|c|c|c|}
		\hline
			\bf{Station} & \bf{Pixel value} & \bf{Clear days} & \bf{Cloudy days} & \bf{Predicted clear days} & \bf{Cloud suit.} \\
		\hline 
			Strathgordon (Tas) & 118 & 16.1 & 211.4 & 28 & 0.15 \\ \hline
			Coonabarabran (NSW) & 96 & 146.3 & 94.5 & 118 & 0.52 \\ \hline
			Alice Springs (NT) & 76 & 200.2 & 63.1 & 193 & 0.85 \\ \hline
			Spring Creek (Vic) & 103 & 79.9 & 159.5 & 87 & 0.40 \\ \hline
			Georgetown (Qld) & 80 & 164.6 & 55.4 & 177 & 0.78 \\ \hline
			Arkaroola (SA) & 81 & 181.8 & 62.4 & 173 & 0.77 \\ \hline
			Wittenoom (WA) & 74 & 183.4 & 64.8 & 200 & 0.88 \\ \hline
		\end{tabular}
			\caption{Number of clear and cloudy days per year at selected BoM sites, with regression prediction for number of clear days, and cloud cover suitability from Equation \ref{NormCloud}.}
		\label{tab:ClearCloudy}
\end{table*}

We perform a geographic analysis to compare the proposed sites to the sites currently hosting research-class telescopes: the Perth Observatory at Bickley (W.A.), the Gingin Observatory (W.A.), Canopus Hill (Tas.), Mt Stromlo (A.C.T.) (currently out of commission), and the Siding Spring Observatory (N.S.W.).

\section{Geographic Analysis}
\subsection{Background}
While in Section 2 we discussed the importance of meteorology in choosing a telescope site, there are other concerns, as described by \citet{Ardeberg}, whose second and third sets of siting criteria are geographic, being the site's altitude (ideally above an atmospheric inversion layer), topography, temperature, temperature stability and atmospheric turbulence, which require on-site measurements for accuracy, air pollution, light pollution and seismic activity. We might naturally add to this list flood, fire and cyclone risk, land availability for the task, and environmental impact assessment concerns. We may also be interested in the fourth set of siting criteria which relate to convenience -- site accessibility and availability of water.\\
We perform a low-level geographic analysis across Australia, combining geographic factors with meteorological factors to describe suitabilities. We may then consider each proposed site in detail \citep{Hotan}.\\

All our site requirements have a spatial component, and thus geographic analysis is appropriate. In this study, we focus on the simple combination of elevation with cloud cover using a Geographic Information System (GIS) \citep{Hotan}. GIS analyses are used in a range of fields. We are interested in gaining information by linking remotely sensed data to other datasets with spatial information that is important to us, in particular, a Digital Elevation Model (DEM). 

\subsubsection{Multi-Criteria Decision Analysis}
One important aspect of GIS is the analysis method and science of Multi-Criteria Decision Analysis (MCDA) \citep{MCDAsotas, Longley, Malc}. MCDA is a process by which a number of parameters affecting the goodness of a location for a task are identified, relative importance of these criteria determined using some decision metric, and the normalised variables combined by some algorithm according to the task and a final value for suitability is obtained.\\ 

A common algorithm is the weighted sum approach where each criterion is multiplied by its weight, each weight $\in (0,1)$ and $\sum w_i = 1$.\\
The weights are obtained via a decision making metric \citep{Clemen, MCDAsotas, Maxwell} that allows the user to determine which of the variables are important and their relative importance in the system. In this study, we have the situation where we cannot \textit{a priori} know what the best weightings for cloud cover and elevation are, so we must try a few different values.\\
A general suitability involving both $k$ Boolean and $l$ continuous criteria, may be given by the rule
\begin{equation}
\label{contBoolSuit}
	suitability = \prod_{a=1}^k criteria_a \times \sum_{b=1}^l (weight_b \times criteria_b).
\end{equation}

\subsection{Using MCDA to choose locations}
In order to perform a geographic analysis to find an appropriate site for a telescope, it is reasonable to use MCDA. The major focus of this study is in finding general regions in which telescope siting is likely to be productive, so we do not focus in detail on the lower-level criteria.\\

We are interested in the rate of cloud cover; and to a low order, we can consider elevation (altitude above sea level) to be a proxy for seeing conditions \citep{Racine}. 
Greater elevation means less atmosphere for light to pass through and potentially different and more stable air columns than at sea level. In reality the amount of atmosphere above a site is more accurately gauged by air pressure. 
We have used the model of \citet{StdAtmos} Equation 33a to check the effect of switching to a pressure based metric and find no significant difference in our final suitability numbers. 

\subsubsection{Sources of data}
The data used in this analysis is that created in Section 2, the high--resolution averaged cloud cover over Australia between June 2008 and January 2010. All geographic data used in this study were obtained from Geoscience Australia's free data downloads\footnote{https://www.ga.gov.au/products/}. The Digital Elevation Model (DEM) used was the 9 arcsecond DEM, which relates to a grid size of roughly 250m square. This is significantly higher resolution than the cloud cover data.\\
Further datasets were downloaded from Geoscience Australia as required. Some data were manually created, such as the layer showing present and proposed telescope locations.

\subsubsection{MCDA method}                                                                                       
In order to combine ``layers'' in our GIS, we need to normalise our criteria so that we are able to compare them. It is important to note at this point that this study is predicated on wishing to build a telescope \textit{in Australia}, and thus this normalisation to create suitabilities only applies within Australia. This allows for relative comparison of the goodness of sites over the region considered.\\

The DEM gives us altitude values in a grid across Australia, which we scale so that the altitude suitability is given by
\begin{equation}
\label{NormElevation}
altitude_{suit} = \frac{elevation}{Max(elevation)},
\end{equation} 
where maximum elevation is $2228m$. We define this criterion's suitability on the basis that higher is always better, although sites in different climates may in practise have equivalent seeing at different altitudes.\\

We normalise cloud cover data such that areas of low cloud cover have high suitability and areas of high cloud cover have low suitability.\\
We scale the cloud cover data to its criterion suitability, while maximising the dynamic range of the data. By inspection, all continental data points take pixel values between approximately 67 (minimal cloud cover) and 127 (maximal cloud cover). Thus we can linearly rescale and normalise this layer (given the linear trend seen in Figure \ref{fig:ClearCloudy}) according to
\begin{equation}
\label{NormCloud}
cloud_{suit} = 1 - \frac{PixelValue(cloud) - 67}{127 - 67}.
\end{equation}

We can combine these two criteria suitability layers together using Equation \ref{contBoolSuit}. First we need to determine what the weighting of each criterion should be. There is no clear answer as to which is more important, high elevation or low cloud cover. Instead we wish to optimise the two. The best approach is determined to be to calculate suitabilities for a number of combinations of criteria weightings, and discuss the implications of each. 

\subsection{Results of MCDA}
\label{MCDAresult}
\begin{figure}[htbp]
	\centering
		\includegraphics[scale=0.25]{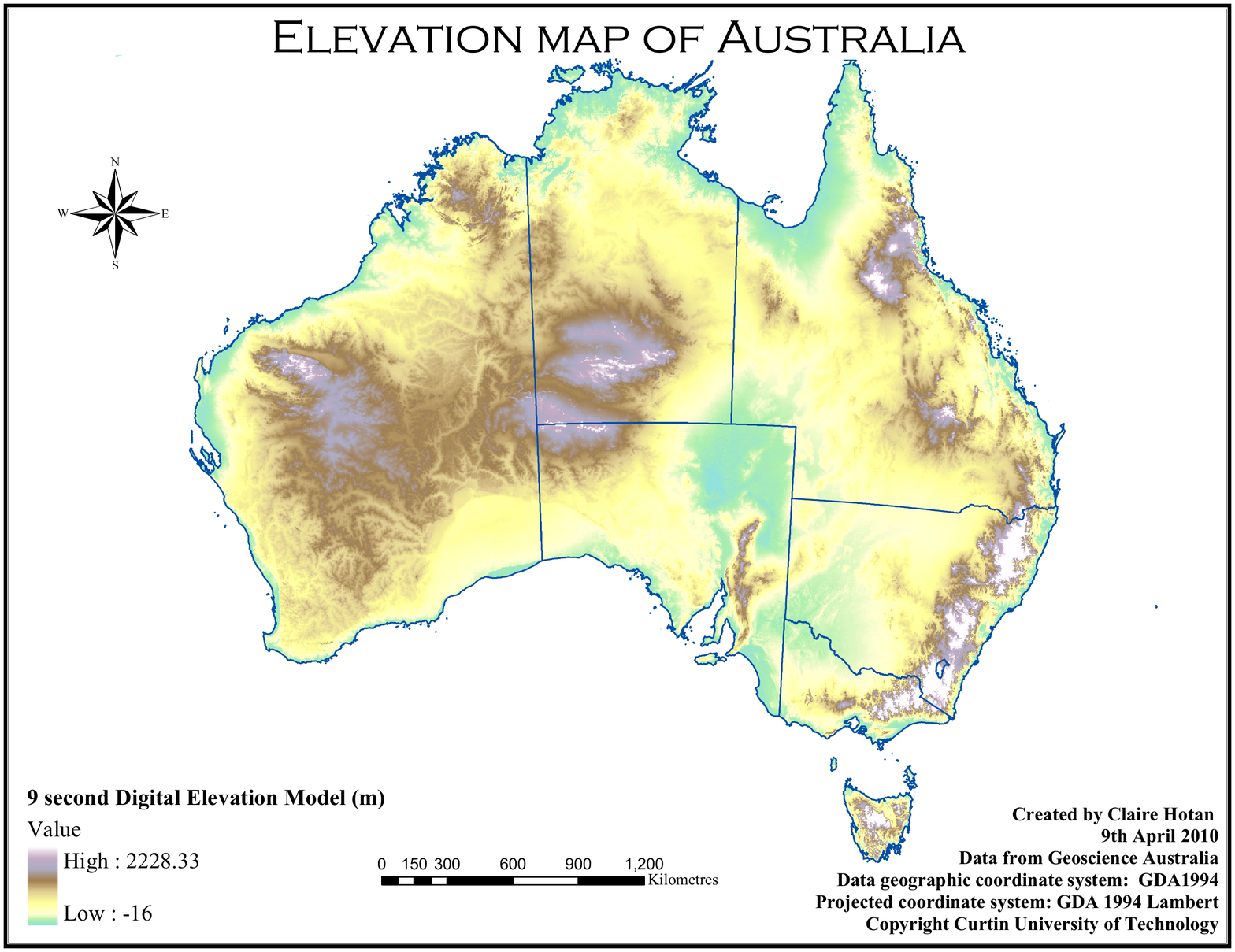}
		\caption{Map showing elevation across Australia at {\raise.17ex\hbox{$\scriptstyle\sim$}}250m resolution.}
	\label{fig:DEM}
\end{figure}

We have the DEM shown in Figure \ref{fig:DEM}. The figure is coloured to 3 standard deviations, meaning that we mostly colour the near--mean detail and set the colour--scale to its maximum or minimum value when we are significantly far from the mean value. 
Intuitively, if we are interested in elevation, we would primarily look for sites along the Great Dividing Range, and also consider parts of Tasmania, Queensland, South Australia, and the inland regions of the Northern Territory and northwest Western Australia.\\

We create a comparable map of cloud cover suitability, which we obtain by rescaling our summed meteorological data as given by Equation \ref{NormCloud}, and shown in Figure \ref{fig:cloudMap}. Note that we resampled the layer to have the same resolution as the DEM by interpolation to crop the cloud cover data accurately to the DEM.\\

We now combine these two layers (shown in Figures \ref{fig:cloudMap} and \ref{fig:DEM}) using the MCDA method given by Equation \ref{contBoolSuit}. We will consider 3 cases. Each case represents a different set of conditions which will be appropriate to a different type of telescope or observing. We always want elevation to be the key factor, as atmospheric clarity will always be important.

\subsubsection{Case 1: Equal weightings}
\begin{figure}[htbp]
	\centering
		\includegraphics[scale=0.25]{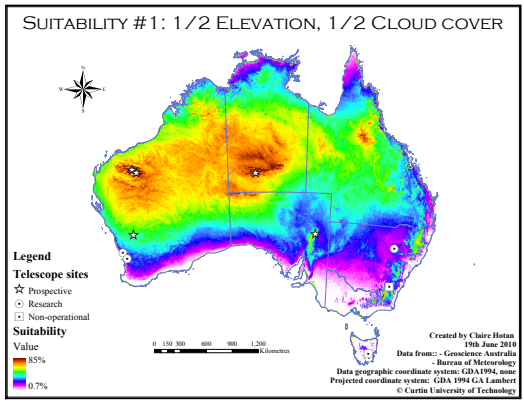}
		\caption{Map showing suitability values across Australia for an ``Equal weightings'' metric.}
	\label{fig:Suit1cont}
\end{figure}

Here we consider low cloud cover and high elevation to be equally important. Such a suitability metric would be appropriate for a telescope on which we value having a lot of observing time, with goodness of atmosphere of no greater importance than the time we can spend observing. 
This metric would be suitable for siting a telescope whose primary purpose is transient follow--up work and Near Earth Object (NEO) detection and tracking. \\
Figure \ref{fig:Suit1cont} shows the result of combining the data in Figures \ref{fig:cloudMap} and \ref{fig:DEM} by multiplying each by 0.5 and adding them together (both are scaled to between 0 and 1).

This map also shows the locations of existing telescopes and candidate sites, to give the reader a feel for the location and approximate suitability of each location under this metric, as do the following maps.

\subsubsection{Case 2: Twice elevation}
\begin{figure}[htbp]
	\centering
		\includegraphics[scale=0.25]{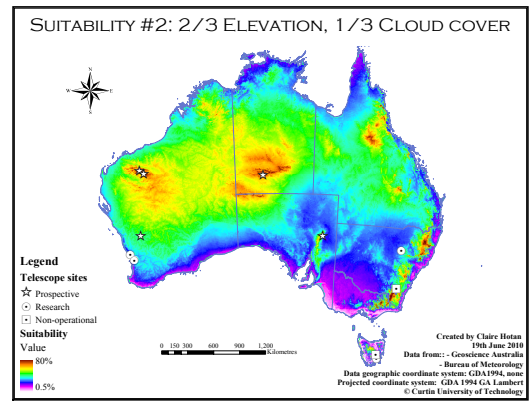}
		\caption{Map showing suitability values across Australia for a ``Two times elevation'' metric.}
	\label{fig:Suit2cont}
\end{figure}

In this case we consider the elevation to be twice as important as cloud cover. This metric would be appropriate for a telescope in which we require fairly good seeing, and much observing time. Extensions to work done at the Australian Astronomical Observatory may be possible with such a telescope, such as surveys like 6dF \citep{Jones} and WiggleZ \citep{Drinkwater}.\\ 
Figure \ref{fig:Suit2cont} shows the result of combining the data in figures \ref{fig:cloudMap} and \ref{fig:DEM} by combining them with a weighting of {\raise.17ex\hbox{$\scriptstyle\sim$}}0.667 on the DEM and {\raise.17ex\hbox{$\scriptstyle\sim$}}0.333 on the cloud cover rate.

\subsubsection{Case 3: 10\% cloud cover}
\begin{figure}[htbp]
	\centering
		\includegraphics[scale=0.25]{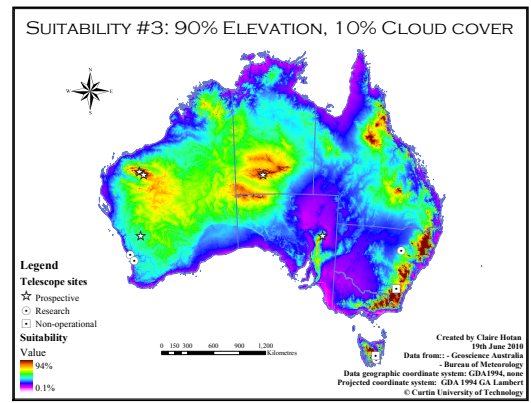}
		\caption{Map showing suitability values across Australia for a ``10\% cloud cover'' metric.}
	\label{fig:Suit3cont}
\end{figure}

In this case we consider looking for a site where we are predominantly concerned with elevation, with low cloud cover being a secondary factor. Such a metric is useful for siting telescopes where astronomical seeing is of prime importance. Such a site would allow very good imaging. A higher elevation site will also generally be superior in the infrared.\\ 
Figure \ref{fig:Suit3cont} shows the result of combining the data in figures \ref{fig:cloudMap} and \ref{fig:DEM} by adding them with a weighting of 0.90 on the DEM and 0.10 on cloud cover rate.

\subsection{Discussion of MCDA results}
In Section \ref{MCDAresult}, note that for the first metric, where elevation and cloud cover are given equal importance, we see that roughly anywhere in the Northwest area of Australia would be suitable, with the Southern and Eastern areas of the country excluded due to their relatively high rates of cloud cover. However when we consider the third case, in which 90\% of the suitability comes from the site's elevation, we find that parts of the Great Dividing Range are considered suitable while large areas in the Northwest of the country are not. This is because there are some relatively high peaks in the this range, so in some cases if the cloud cover in those areas is low enough, we will still rate them as suitable, even though they are likely to have a relatively low rate of photometric observing conditions, similar to and perhaps worse than Siding Spring, N.S.W.. As one might expect, the results of the intermediate case metric produce suitability areas which are similar to both extreme cases.\\

We note that the observatories located at the Siding Spring are in perhaps one of the best locations, within the constraints of being built an accessible distance from major cities. 

In the remainder of this section we discuss the identification of prospective telescope sites based on this analysis, and compare those sites to existing sites, and sites proposed previously by other authors.

\subsubsection{Identifying potential sites}
It is apparent that some areas of Australia appear to perform consistently well over all metrics.\\
If we have a particular science objective, or potential telescope that we wish to site, we may concentrate on the most suitable metric, however in this instance we are looking for typically excellent sites for prospective astronomical observing, then we consider the results produced in Figures \ref{fig:Suit1cont}, \ref{fig:Suit2cont} and \ref{fig:Suit3cont} to find locations or areas which promise to be capable of producing good results for any telescope which may be built. By inspection of these figures, we see consistently suitable areas arising in the ranges in northwest of W.A. and inland N.T., with both ranges having peaks of relatively high elevation. \\
From a meteorological perspective, prevailing mid--latitude winds blow from west to east, meaning that we might expect better atmospheric stability over western mountains than eastern mountains. The particular details of cloud formation and medium--scale meteorology over areas of Australia is beyond the scope of this study, but may be considered a factor which could negatively impact on sites.\\

Thus based on this method of combining average cloud cover data with elevation data, we can propose sites which may in the future be tested for siting optical research telescopes in Australia. These sites are Mt. Bruce and Mt. Meharry in the Hamersley Range (W.A.), and the MacDonnell Ranges in the vicinity of Mereenie (N.T.). These sites have peaks of approximately 1200m (W.A.) and 940m (N.T.). 

\subsubsection{Comparison of proposed sites with previously suggested sites}
The three sites we proposed as GIS--based candidates are now compared with other sites. To do this, we need to ``extract'' the suitability rating from each metric for each site of interest. Table \ref{tab:siteSuits} shows the locations and suitabilities of the sites of interest in this study -- present research sites, those proposed by other studies, and those proposed here. The table also shows the approximate elevation of each site, as well as a prediction of the approximate number of clear days for that site based on a regression of Figure \ref{fig:ClearCloudy} using the pixel value at each site.\\

In Table \ref{tab:siteSuits} it is apparent that most of the currently existing research telescopes are not ideally located. The Siding Spring Observatory fares remarkably well under the first metric of equal weighting of cloud cover and elevation with a suitability of 53\%, considering that our primary reason for wanting to find a new telescope site is the frequent poor observing conditions at this site. Perth Observatory and Gingin Observatory, both in W.A., which are both in a latitude zone with dry summers but very wet winters, perform poorly under this metric, and both are at low elevations. Similarly outperformed in this metric are the defunct Mount Stromlo site (A.C.T.) and both Canopus Hill and the new Bisdee Tier site in Tasmania. \\
All ``proposed'' sites carry similar or higher suitability ratings under the first metric than Siding Spring. Note that we would predict Siding Spring will experience 241 non--cloudy days per year, while the Hamersley Range sites are predicted to experience 319 non--cloudy days. Indeed, \citet{Lee} states that in the period of interest, the AAO was able to observe 64\% of the night time, which is roughly equivalent to 233 non--cloudy days.\\ 

\begin{sidewaystable}[p]
	\centering
	\small
		\begin{tabular}{lllllcllccc}
		\hline
		\bf{Latitude} & \bf{Longitude} & \bf{Status} & \bf{Name} & \bf{State} & \bf{Size} & \bf{Elevation} & \bf{Clear\footnote{Predicted number of clear days based on Figure \ref{fig:ClearCloudy}, where pixel values for each site are manually extracted using ArcGIS.}} & \bf{\textit{Suitability \#1}} & \bf{\textit{Suitability \#2}} & \bf{\textit{Suitability \#3}} \\
		\hline
		-31.2754 & 149.0672 & Research & Siding Spring & NSW & 3.9m & {\raise.17ex\hbox{$\scriptstyle\sim$}}1130m & 114 days & 53\% & 51\% & 48\% \\
		-42.8475 & 147.43296 & Research & Canopus Hill & Tas & 1.0m & {\raise.17ex\hbox{$\scriptstyle\sim$}}260m & 40 days & 20\% & 16\% & 11\% \\
		-32.0073 & 116.1367 & Research & Perth Observatory & WA & 0.6m & {\raise.17ex\hbox{$\scriptstyle\sim$}}390m & 134 days & 41\% & 34\% & 22\% \\
		-31.35598 & 115.713115 & Research & Gingin & WA & 1.0m & {\raise.17ex\hbox{$\scriptstyle\sim$}}50m & 150 days & 38\% & 25\% & 9\% \\
		-35.31903 & 149.00883 & Defunct\footnote{Destroyed by bushfire in 2003 and not fully rebuilt.} & Mount Stromlo & ACT & 1.9m & {\raise.17ex\hbox{$\scriptstyle\sim$}}770m & 95 days & 43\% & 40\% & 35\% \\
		-42.42594 & 147.2885 & Pending\footnote{The University of Tasmania is relocating its primary observatory site from Canopus Hill to Bisdee Tier, this site is currently under construction.} & Bisdee Tier & Tas & 1.3m & {\raise.17ex\hbox{$\scriptstyle\sim$}}600m & 40 days & 27\% & 27\% & 25\% \\
		-29.4675 & 117.300 & Proposed\footnote{Proposed and studied by Biggs and \citet{Walsh}.} & Mt. Singleton & WA & -- & {\raise.17ex\hbox{$\scriptstyle\sim$}}670m & 165 days & 52\% & 42\% & 29\% \\
		-30.3082 & 139.3381 & Proposed\footnote{Proposed by \citet{Coops} and also studied by \citet{Hogg65} and \citet{Wood}.} & Freeling Heights & SA & -- & {\raise.17ex\hbox{$\scriptstyle\sim$}}940m & 177 days & 48\% & 37\% & 21\% \\
		-22.608 & 118.144 & Proposed\footnote{Proposed by KG.} & Mt. Bruce & WA & -- & {\raise.17ex\hbox{$\scriptstyle\sim$}}1200m & 197 days & 71\% & 65\% & 57\% \\
		-22.980 & 118.588 & Proposed\footnote{Proposed in this study.} & Mt. Meharry & WA & -- & {\raise.17ex\hbox{$\scriptstyle\sim$}}1200m & 201 days & 72\% & 66\% & 58\% \\
		-23.88611 & 132.20 & Proposed$^g$ & Mereenie & NT & -- & {\raise.17ex\hbox{$\scriptstyle\sim$}}950m & 193 days & 61\% & 53\% & 40\% \\
		\end{tabular}
			\caption{Locations, predicted clear days and suitability values for present and proposed telescope sites.}
		\label{tab:siteSuits}
\end{sidewaystable}

The Siding Spring Observatory telescopes are typically used for galaxy redshift surveys and similar science. Thus we wish to be in the regime around the second metric, where elevation carries more weight than cloud cover, but cloud cover remains an important consideration. In this metric, Siding Spring is given a suitability value of 51\%. Note that 100\% suitability would require no cloud cover at maximum elevation which is highly unlikely. Three of the five proposed sites have equal or better suitabilities under the second metric than Siding Spring. Thus it appears reasonable to expect that if we wished to build a new telescope to perform similar work to that done by the telescopes at Siding Spring, we would do well to consider siting it at one of these 3 suggested sites.\\

In the third metric, we place the greatest importance in elevation, as an indication of good seeing when conditions are photometric, we find that both Siding Spring and Mount Stromlo have values of around 48\% and 35\% respectively, while the other active research telescopes are much lower. We note that it would appear that the relocation of the University of Tasmania's primary observatory from Canopus Hill to Bisdee Tier should more than double the suitability of their site under this metric, although resolution of less than 1 arcsecond remains unlikely \citep{Cole}. The two previously proposed and tested sites, Mount Singleton (W.A.) and Freeling Heights (S.A.) both appear to be good sites, but by this metric do not hold any advantage over existing sites. In the case of Freeling Heights this appears to be in contradiction to the findings of \citet{Wood}. The proposed sites in northwest Western Australia appear very well suited to this metric.\\ 

An examination of each of the identified sites, including accessibility and possible light pollution, is presented in CH's dissertation \citep{Hotan}.

\subsubsection{Prospective sites for a new telescope}
The mountains in the Hamersley Range in the Pilbara region of Western Australia have been identified in a number of previous studies based on meteorological analyses of cloud cover patterns, but typically discounted due to remoteness. 
In recent years astronomers have started to consider sites where accessibility concerns are not considered important, wishing instead to gain the best sites available, such as those in Chil\'{e}, and Antarctica \citep{Burton}.\\
Under the assumption that we should be more interested in the goodness of a site than the practicalities of building on or accessing that site, KG performed a meteorological analysis, concluding that Mount Bruce could be potentially the best site in Australia for a future optical telescope in this country. It is worth noting that the studies mentioned previously, and the current study, are performed with data from different epochs, so the fact that this area is repeatedly preferred suggests that it is less likely to be due to chance climatic variations.\\
Thus based on the information presented in Table \ref{tab:siteSuits} we might suggest that both of the two mountains identified in the Hamersley Range are good candidates and should be considered seriously, with site testing performed at these sites. We comment though that the Northern part of Australia typically experiences higher rates of cyclones and lightning strikes\footnote{See e.g. lightning maps on  http://thunder.msfc.nasa.\\gov/data/OTDsummaries/gifs/1999\_YTD\_world.gif.}.\\

Practical considerations need to be taken into account, in this case, the consideration in the third group of criteria suggested by \citet{Ardeberg}, of artificial light pollution. 
Western Australia is very mineral--rich, and there are many mines throughout the outback regions. One such very large mine is located on the side of Mount Bruce, and so despite the height of the mountain, it appears likely that sky quality conditions may be adversely affected on this site. Thus it is considered pertinent to also investigate the nearby Mount Meharry, which is located within the Karijini National Park, and is therefore somewhat shielded from such light pollution. By the model of \citet{Racine} we might expect sub-arcsecond median seeing values a little over 0.9\textquotedblright at these sites for a telescope elevated 4m above the ground.\\ 
Although these two mountains are geographically close together, they differ in topography. This may lead to differing air-flow patterns across their peaks, which could produce different observing conditions at each site.\\ 

Based solely on our suitability calculations presented in Table \ref{tab:siteSuits}, it appears that sites in the MacDonnell Ranges may be of similarly high suitability ratings to those in the Hamersley Range in Western Australia, and may have higher elevations and suitabilities, as the highest point in the MacDonnell Ranges is Mount Zell at $1531m$. Intuitively we would expect significant diurnal convection currents this far inland, and would not expect stable air columns. Nevertheless, based only on the suitability values computed, without access to good atmospheric models for each area, we can not necessarily recommend the Hamersley Range sites over those potentially available in the MacDonnell Ranges. Thus a site such as that suggested here should also be physically tested to ascertain the atmospheric behaviour in this region.

\subsection{Future extensions to this work}
The initial and important extension to this work would be performing site tests to obtain empirical values for the astronomical seeing at the proposed sites. 
We should obtain comparison values from operational observatories.\\

Further geographic analysis is also necessary prior to making proposals to government or universities in terms of land acquisition and logistics. Geoscience Australia provides appropriate datasets for this task. \\
We wish to know that the site is geologically stable. 
In terms of potential light pollution it will also be important to try to build in a protected area, and that no significant mineral deposits which could become mining targets in the future lie within the ``viewshed'' of the proposed site.\\ 

We should ideally also extend the meteorological work to cover a longer time range. It would also be preferable to obtain precipitable water vapour data, and perform an equivalent analysis. Ideally a number of climate models should also be studied to look at present and recent past meteorological suitability of a site, and whether we would predict the goodness of the site will improve or worsen over the medium term of its operation. \citet{SEACI}, for example, study diminishing rainfall in South East Australia, and \citet{IOCI} study the science of increased rainfall and extreme weather events in northwest Western Australia.

\section{Conclusion}
In this study we have investigated the prospective siting for any potential optical telescope to be built within Australia. We obtained IR cloud cover data from the Bureau of Meteorology, which we combined to find trends in total cloud cover over the 18 month period June 2008 -- January 2010. This analysis indicated areas of low cloud cover over the northwest of Western Australia, extending inland to the Northern Territory over the MacDonnell Ranges.\\
We then imported the total cloud cover image into GIS software, and combined it with a digital elevation model of Australia to find areas suitable for various types of telescopes, using three different metrics of weighting the importance of elevation (to minimise atmospheric disturbance and extinction) and cloud cover. While different weightings produced different results, the mountains in the Hamersley Range of northwest Western Australia consistently performed well in all the metrics. They have both good elevations by Australian standards and very low cloud cover rates.

\section*{Acknowledgments} 
This work has been carried out as part of CH's Masters dissertation research under the supervision of ST and other staff at ICRAR, with advice from KG. Thanks to those who helped proofread this manuscript.

\end{document}